\documentclass[aps,prl,preprint,groupedaddress,linenumbers]{revtex4}
 \usepackage{graphicx}
 \usepackage{dcolumn}
 \usepackage{bm}
 \usepackage{amsmath}
 \usepackage{color}
 \usepackage{amssymb}
 \usepackage{mathrsfs}
 \usepackage{MnSymbol}%
 \usepackage{wasysym}%
 \usepackage{gensymb}
 \usepackage{framed,color}
 \usepackage{enumitem}

\newenvironment{sciabstract}{%
	\begin{quote} \bf}
	{\end{quote}}

\begin{document}


\title{Unequivocal Determination of Spin-Triplet Superconductivity Using Composite Rings}


\author{Xiaoying Xu,$^{1}$ Yufan Li,$^{1,2}$ C. L. Chien$^{1,3}$}
\affiliation{$^1$William H. Miller III Department of Physics and Astronomy, Johns Hopkins University, Baltimore, MD 21218, USA}
\affiliation{$^2$Department of Physics, The Chinese University of Hong Kong, Shatin, Hong Kong SAR, China}
\affiliation{$^3$Department of Physics, National Taiwan University, Taipei, 10617, Taiwan}

\date{\today}

\baselineskip24pt

\maketitle

\begin{sciabstract}
	Phase-sensitive measurements on a composite ring made of a superconductor of interest connected by a known singlet $s$-wave superconductor can unambiguously determine its pairing symmetry. In composite rings with epitaxial $\beta-$Bi$_2$Pd and $s$-wave Nb, we have observed half-integer quantum flux when Nb is connected to the opposite crystalline ends of $\beta-$Bi$_2$Pd and integer-quantum flux when Nb is connected to the same crystalline ends of $\beta-$Bi$_2$Pd. These findings provide unequivocal evidence of odd-parity pairing state of the triplet superconductor $\beta-$Bi$_2$Pd. 

\end{sciabstract}

\begin{text}

Superconductivity is the result of condensation of Cooper pairs, where the pairing of two electrons can form either spin singlet with spin 0, or spin triplet with spin 1. 
To satisfy Fermi statistics, the spatial part of the pair's wave function must be of even parity for spin-singlet, and of odd parity for spin-triplet pairing states. 
The known superconductors (SCs) are overwhelmingly singlet-pairing, including the conventional $s$-wave SCs and the high-$T_c$ $d$-wave cuprates. 
The very rare triplet pairing state has only been confirmed in $^3$He superfluid \cite{He3_book}. 
The decades-long quest of searching for intrinsic triplet SCs is currently met with renewed interest as it has been shown that triplet pairing states generally lead to topological superconductivity, essential for realizing Majorana fermions with non-Abelian brading statistics that facilitates noise-resilient topological quantum computing \cite{qi_topological_2011,alicea_new_2012,sato_topological_2017}. 

By the nature of their pairing states, one can decisively distinguish between singlet and triplet SCs by the parity symmetry of its wave function via phase-sensitive experiments \cite{geshkenbein_vortices_1987,tsuei_pairing_2000}. 
It is first proposed by Geshkenbein, Larkin, and Barone (GLB) that a signature half-integer quantum flux (HQF) can be observed in a composite ring structure, consisting of a triplet $p$-wave SC connected by a singlet $s$-wave SC at the opposite ends \cite{geshkenbein_vortices_1987}. 
It is a straightforward testament of the odd parity for triplet pairing as the sign of the gap function reverses upon the inversion of the momentum, or $\Delta_k = -\Delta_{-k}$. 
This sign reversal is experienced at the pair of $s-p$ junctions with opposite normal directions, contributing a $\pi$ phase shift that leads to fluxoid quantization on half-integer quantum numbers; $\Phi^{\prime} = (n+1/2)\Phi_0$ where $n$ is an integer. 

The first experimental demonstrations of the phase-sensitive methods are in fact a variant of the GLB proposal to reveal the $d$-wave pairing symmetry in high-$T_c$ cuprates \cite{Sigrist1992}. 
Two orthogonally oriented crystal planes of a single crystal YBa$_2$Cu$_3$O$_7$ are connected with a Pb thin film, constituting a corner SQUID / corner junction geometry \cite{wollman_experimental_1993, Wollman1995}. 
The sign change of the gap function under 90$^{\circ}$ rotation results a $\pi$ phase shift in the Josephson current-phase relation, thus revealing the $d_{x^2-y^2}$ gap structure.

The GLB experiment, in its original design and for the proposed purpose of detecting odd parity symmetry, was carried out in Sr$_2$RuO$_4$-Au$_{0.5}$In$_{0.5}$ SQUID device to examine the proposed triplet $p$-wave pairing of Sr$_2$RuO$_4$ \cite{nelson_odd-parity_2004}. 
In all the aforementioned studies, the samples of interest are bulk single crystals. 
An inherent challenge with this approach is rooted in the small value of the flux quantum $\Phi_0$, approximately 20.7~Gauss-($\mu$m)$^2$. 
The large sizes of devices when involving bulk specimen, generally on the 100 $\mu$m scale, lead to small flux quantization periods in terms of magnetic field on the order of mGauss \cite{wollman_experimental_1993,nelson_odd-parity_2004}. 
Extrapolation to identify the zero magnetic field is required for establishing the HQF, thus inevitably leaves room for ambiguity \cite{tsuei_pairing_2000,Mackenzie2017}.  

We instead employ a planar composite ring structure, comprising of thin films of epitaxial $\beta-$Bi$_2$Pd, a triplet $p$-wave SC candidate and Nb, an $s$-wave SC. 
The thin film specimen has the advantage of allowing the device to be defined by electron-beam (E-beam) lithography. 
In this study, we fabricated $\mu$m-sized composite ring devices with the $\Phi_0$-period on the order of 10~Oe in terms of magnetic field. 
The presence of HQF can be straightforwardly determined from the Little-Parks effect \cite{Li2019,xu_spin-triplet_2020}. 
The result shows clear-cut evidences for odd-parity paring in $\beta-$Bi$_2$Pd. 
Indeed, by resolving the parity symmetry of the superconducting gap, the composite ring structure can unequivocally reveal singlet or triplet pairing of any superconductors. 
We show that the HQF can be achieved on demand, only requiring appropriate design geometry. 
The accessible HQF readily integrable with 2D lithography technique opens door for applications of zero-field flux qubits \cite{Li2019}. 

We first synthesized epitaxial (001)~$\beta-$Bi$_2$Pd films on (001)~SrTiO$_3$ substrates with magnetron sputtering, exploiting the fact that the $ab$-plane of $\beta-$Bi$_2$Pd with a tetragonal lattice with $a=b=$3.36~$\AA$ and $c=$12.98~$\AA$ can be epitaxially grown on the (001)~SrTiO$_3$ surface \cite{Lv2017,Li2020}. 
The x-ray diffraction pole-figure measurements of the epitaxial films establishes the epitaxial relationship of [100]~$\beta-$Bi$_2$Pd~$\parallel$~[100]~SrTiO$_3$ with details in the Supplementary Materials. 
Composite ring devices are patterned by E-beam lithography to the sub-$\mu$m level, consisting of a segment of epitaxial $\beta-$Bi$_2$Pd and the remainder of polycrystalline Nb. 
We explored two classes of geometries. 
The first geometry is the original GLB proposal \cite{geshkenbein_vortices_1987}. 
A typical device is shown in Fig.~1e as the scanning electron microscopic image of GLB-A, where a straight segment of epitaxial $\beta-$Bi$_2$Pd is connected by Nb at the \textit{opposite} ends. 
HQF is expected for the odd-parity pairing state in this geometry.  
In the second geometry, the \textit{same} side of a crystal plane face of $\beta-$Bi$_2$Pd is connected by Nb, as the device SS-B shown in Fig.~1f. 
Conventional integer-quantum flux (IQF) is expected for this geometry, regardless of the pairing state. 
The temperature dependence of a typical ring device shows two sharp transitions at about 3~K and 8~K due to the superconducting transitions of $\beta-$Bi$_2$Pd and Nb respectively, as shown in Fig.~1g. 

The Little-Parks effect reveals the oscillation of the superconducting critical temperature $T_c$ as a function of the applied magnetic field threading through the ring, reflecting the periodic variation of the free energy as a result of fluxoid quantization \cite{little_observation_1962}. 
Experimentally, one observes the oscillation of the resistance of the ring at a fixed temperature below but close to $T_c$. 
For the conventional integer flux quantization, i.e., $\Phi^{\prime} = n\Phi_0$, the resistance minima occurs at integer quantum numbers, as shown in Fig.~1d. 
For HQF, the resistance minima occurs at half-integer quantum numbers. 
To avoid artifacts such as trapped vortex, the sample is always cooled down in zero field from 10~K before each measurement. 
The result is always confirmed by sweeping the field in both directions. 
The Little-Parks effect of normalized resistance as a function of perpendicularly applied magnetic field in GLB-A at 2.7~K is demonstrated in Fig.~2a. 
The measured oscillation period of 4.25~Oe is consistent with the estimated period of about 3.1~Oe from the intended dimension of 4.2~$\mu$m~$\times$~1.6~$\mu$m. 
For the Little-Parks effect, an aperiodic background originating from, among others, a slight field misalignment \cite{tinkham_consequences_1964} and device geometry \cite{Moshchalkov1995} generally appears with the $\Phi_0$-periodic oscillations. 
After the curved background is removed from the raw data in the upper panel, pure Little-Parks oscillation is plotted in the lower panel, which shows the resistance minima at fields corresponding to half-integer flux of $\Phi = (n+1/2)\Phi_0$. 
The determination of the zero-field position is straightforward because of the substantial field values of several to several tens of Oe, and that can be further aided by multiple features of the Little-Parks oscillations, such as the symmetry of the curved background, and the symmetrical glitches near $\pm$5.5~$\Phi_0$. 
The observation of HQF, the telltale sign of odd-parity pairing, conclusively shows that $\beta-$Bi$_2$Pd is a spin-triplet SC. 

In contrast, the SS-B device demonstrates only the ordinary IQF as would be expected for any doubly-connected superconducting devices. 
The Little-Parks effect shows an oscillation period of 31.0~Oe, in close agreement with the 27.0~Oe as expected from the ring size of 1.1~$\mu$m~$\times$~0.7~$\mu$m. 
The resistance minima occur only at integer flux of $\Phi = n\Phi_0$. 
The determination of the zero-field position again can be aided by other features such as the symmetry of the background as well as the symmetrical pair of imperfections near $\pm$7~$\Phi_0$. 

The core reason of the profoundly different fluxoid quantization observed in GLB-A and SS-B is the slight but essential difference in the design geometries. 
The normal directions of the contacting crystal plane surfaces at the junctions are parallel for SS-B, where the gap function remains the same, as opposed to the antiparallel normal directions for GLB-A, where the gap functions have opposite signs. 
The absence of HQF in SS-B rules out alternative interpretations of the HQF observed in GLB-A. 
Indeed, the HQF could only originate from odd-parity symmetry. 

The Little-Parks effect can be observed within a temperature window in the vicinity of the superconducting phase transition of the ring component with the lowest $T_c$, in our case the $\beta-$Bi$_2$Pd segment. 
At the lower bound of the temperature window, the onset of the Little-Parks oscillations occurs when the $\beta-$Bi$_2$Pd segment starts to develop finite resistance. 
At the upper bound of the temperature window, the phase coherence is lost because much of the $\beta-$Bi$_2$Pd segment is in the normal state, and the Little-Parks oscillations fade out. 
Below but close to the upper bound, there may exist a region where $\beta-$Bi$_2$Pd loses the characteristics of its own odd-parity superconductivity, but relies on the proximity effect of the contacting Nb to sustain the Little-Parks effect. 
In such cases, the composite ring would behave as an even-parity singlet SC, regardless of the GLB or the SS geometry. 
Indeed, we are able to observe in the GLB devices the transitions from HQF at lower temperatures, to IQF at higher temperatures. 
Fig.~3a shows the representative results obtained from GLB-A. 
The oscillations first emerge at 2.6~K with the signature of HQF. 
The resistance minima occur at the half-integer flux $\Phi = (n+1/2)\Phi_0$ for 2.6~K and 2.7~K. 
At 2.75~K, however, the HQF signature persists only for $|n| \geq 3$. 
For $|n| < 3$, a transitional behavior is observed. 
Minima may be found in both integers $n$ or half integers $n \pm \frac{1}{2}$. 
At 2.8~K, the conventional IQF is established for $|n| \leq 4$, and the transition region is pushed towards higher fields, as the oscillations for $|n| > 4$ nearly washed out. 
The IQF completely takes over at 2.9~K. 
At this point, the superconductivity in $\beta-$Bi$_2$Pd segment is dying out rapidly, and no Little-Parks effect can be observed at or above 3.0~K. 
In contrast, the SS-B device demonstrates only IQF throughout the temperature window (1.8~K - 3.5~K) as shown in Fig.~3b, with no signs of HQF. 

The HQF-IQF transition has been generally observed in \textit{all} our GLB composite rings, as summarized in the Supplementary Materials, always with the ascending temperature. 
The HQF and the IQF represent the Little-Parks oscillations with and without the $\pi$ phase shift, respectively. 
Having the occurrence of both in one experiment succinctly lifts any doubts that the $\pi$ phase shift as the signature of HQF has been correctly identified. 
The transition also indicates that the $\pi$ phase shift originated from the odd-parity pairing, which vanishes when the superconductivity of $\beta-$Bi$_2$Pd is suppressed at the highest temperatures. 
It reaffirms that the odd parity is closely related to the intrinsic superconducting phase of $\beta-$Bi$_2$Pd. 
An interesting observation can be made in GLB-A, when IQF is first developed at the low fields at 2.75~K and then pushes towards higher fields at 2.8~K. 
It appears to suggest that a finite field helps to stabilize the triplet pairing state, while it already gives way to the even-parity singlet pairing, induced by proximity effect with Nb, near the zero field. 
This, on the face value, is consistent with the expectation of the spin-parallel configurations of the spin-triplet states. 
However, this field dependence is yet to be confirmed in other GLB devices, as the transitions observed therein do not manifest sufficient fine details to resolve it. 

Another intriguing question is whether the odd-parity gap structure has any angular dependence with respect to the crystalline orientations of $\beta-$Bi$_2$Pd. 
The $\beta-$Bi$_2$Pd segment of GLB-A is patterned along the [100] direction of epitaxial $\beta-$Bi$_2$Pd. 
For a tetragonal lattice, it is equivalent to the [010] direction. 
We pattern a composite ring GLB-C with a 45$^{\circ}$ rotation, as shown in Fig.~4a. 
The $\beta-$Bi$_2$Pd segment is orientated along the [110] direction. 
Despite the new orientation, GLB-C also exhibits HQF at lower temperatures, with a transition to IQF at higher temperatures above 3~K. 
Applying the symmetry of the tetragonal structure, this suggests that the same odd-parity gap exists along the [110] and [1-10] directions, as well as the [100] and [010] directions. 
With no clear angular dependence 45 degrees apart, the result strongly suggests that the odd-parity gap is fully open in the $ab$ plane. 
Two likely $p$-wave configurations consistent with this observation are: 
(a) the ABM state \cite{ABM} of the form $|\Delta|e^{i\phi}sin\theta$, expressed in spherical coordinates with $z$-axis along the tetragonal $c$-axis, first realized in A-phase of He$^3$; 
and (b) the BW state \cite{BW}, where the gap function has constant value but changes sign upon inversion as first confirmed as the B-phase of He$^3$. 
Both configurations are symmetrically allowed for a tetragonal lattice structure \cite{RMP_SRO}.  
On the other hand, thermodynamic property measurements of bulk crystals often show fully gapped structure \cite{herrera_magnetic_2015,kacmarcik_single-gap_2016,muSR_fullygapped,HuiqiuNodeless}. 
Reconciling with these experimental results, the BW state may be a more likely candidate. 
Further experiments are required to fully map out the superconducting gap structure of $\beta-$Bi$_2$Pd. 

Our previous experiment shows that HQFs can be observed in polycrystalline $\beta-$Bi$_2$Pd rings \cite{Li2019}.
A polycrystalline ring of a triplet SC may demonstrate either HQF or IQF with about equal chances, and there are no inter-transitions between the two states. 
For the composite rings with epitaxial $\beta-$Bi$_2$Pd, the only deciding factor for HQF or IQF is the device design. 
HQFs can be produced on demand, so long as the device geometry follows the GLB design of an epitaxial triplet SC connected by a singlet SC. 
The HQF devices, made by thin film deposition and standard lithography, are readily compatible with applications. 
This opens door to applications of HQF in flux qubits \cite{Li2019}. 
For conventional flux qubits comprising only singlet SCs, the degeneracy point must be achieved by applying a magnetic field of a precise value. 
Scaling up the number of flux qubits on the same chip is complicated by the requirement of a bias magnetic field. 
For the composite rings with the GLB design, the degeneracy is achieved at the zero magnetic field. 
Lifting the requirement of a bias field offers essential advantages such as better scalability, suppressed flux noise, \textit{etc}. for incorporating topological SCs into viable quantum information applications, a first step towards realizing the promised fault-tolerant features of topological superconductivity.  


In summary, the odd parity of spin-triplet pairing state in $\beta-$Bi$_2$Pd has been unequivocally verified by HQFs observed in composite ring devices comprising of epitaxial $\beta-$Bi$_2$Pd and $s$-wave SC Nb. 
HQF can be obtained on demand by the geometrical design of the composite ring. 
Transitions between HQF and IQF have been observed when the superconductivity of $\beta-$Bi$_2$Pd is suppressed at higher temperatures and superseded by that of the neighboring Nb. 
No angular dependence of the odd-parity gap has been observed along the $<100>$ and $<110>$ directions, implying that the gap may be isotropic in value and opposite in sign upon inversion within the $ab$-plane. 
Most advantageously, a single GLB-design composite ring can unequivocally reveal the singlet or the triplet nature of an unknown SC.

\end{text}


\begin{thebibliography}{25}
	\expandafter\ifx\csname natexlab\endcsname\relax\def\natexlab#1{#1}\fi
	\expandafter\ifx\csname bibnamefont\endcsname\relax
	\def\bibnamefont#1{#1}\fi
	\expandafter\ifx\csname bibfnamefont\endcsname\relax
	\def\bibfnamefont#1{#1}\fi
	\expandafter\ifx\csname citenamefont\endcsname\relax
	\def\citenamefont#1{#1}\fi
	\expandafter\ifx\csname url\endcsname\relax
	\def\url#1{\texttt{#1}}\fi
	\expandafter\ifx\csname urlprefix\endcsname\relax\def\urlprefix{URL }\fi
	\providecommand{\bibinfo}[2]{#2}
	\providecommand{\eprint}[2][]{\url{#2}}
	
	\bibitem[{\citenamefont{Vollhardt and Wolfle}(1990)}]{He3_book}
	\bibinfo{author}{\bibfnamefont{D.}~\bibnamefont{Vollhardt}} \bibnamefont{and}
	\bibinfo{author}{\bibfnamefont{P.}~\bibnamefont{Wolfle}},
	\emph{\bibinfo{title}{The superfluid phases of helium 3}}
	(\bibinfo{publisher}{Taylor \& Francis}, \bibinfo{year}{1990}), ISBN
	\bibinfo{isbn}{0850664128}, \urlprefix\url{https://doi.org/10.1201/b12808}.
	
	\bibitem[{\citenamefont{Qi and Zhang}(2011)}]{qi_topological_2011}
	\bibinfo{author}{\bibfnamefont{X.-L.} \bibnamefont{Qi}} \bibnamefont{and}
	\bibinfo{author}{\bibfnamefont{S.-C.} \bibnamefont{Zhang}},
	\bibinfo{journal}{Rev. Mod. Phys.} \textbf{\bibinfo{volume}{83}},
	\bibinfo{pages}{1057} (\bibinfo{year}{2011}),
	\urlprefix\url{https://link.aps.org/doi/10.1103/RevModPhys.83.1057}.
	
	\bibitem[{\citenamefont{Alicea}(2012)}]{alicea_new_2012}
	\bibinfo{author}{\bibfnamefont{J.}~\bibnamefont{Alicea}},
	\bibinfo{journal}{Rep. Prog. Phys.} \textbf{\bibinfo{volume}{75}},
	\bibinfo{pages}{076501} (\bibinfo{year}{2012}), ISSN
	\bibinfo{issn}{0034-4885},
	\urlprefix\url{http://stacks.iop.org/0034-4885/75/i=7/a=076501}.
	
	\bibitem[{\citenamefont{Sato and Ando}(2017)}]{sato_topological_2017}
	\bibinfo{author}{\bibfnamefont{M.}~\bibnamefont{Sato}} \bibnamefont{and}
	\bibinfo{author}{\bibfnamefont{Y.}~\bibnamefont{Ando}},
	\bibinfo{journal}{Reports on Progress in Physics}
	\textbf{\bibinfo{volume}{80}}, \bibinfo{pages}{076501}
	(\bibinfo{year}{2017}), ISSN \bibinfo{issn}{0034-4885, 1361-6633},
	\urlprefix\url{http://stacks.iop.org/0034-4885/80/i=7/a=076501?key=crossref.db5f5db067ad36cf0669f5cbf8ccb916}.
	
	\bibitem[{\citenamefont{Geshkenbein et~al.}(1987)\citenamefont{Geshkenbein,
			Larkin, and Barone}}]{geshkenbein_vortices_1987}
	\bibinfo{author}{\bibfnamefont{V.~B.} \bibnamefont{Geshkenbein}},
	\bibinfo{author}{\bibfnamefont{A.~I.} \bibnamefont{Larkin}},
	\bibnamefont{and} \bibinfo{author}{\bibfnamefont{A.}~\bibnamefont{Barone}},
	\bibinfo{journal}{Phys. Rev. B} \textbf{\bibinfo{volume}{36}},
	\bibinfo{pages}{235} (\bibinfo{year}{1987}),
	\urlprefix\url{https://link.aps.org/doi/10.1103/PhysRevB.36.235}.
	
	\bibitem[{\citenamefont{Tsuei and Kirtley}(2000)}]{tsuei_pairing_2000}
	\bibinfo{author}{\bibfnamefont{C.~C.} \bibnamefont{Tsuei}} \bibnamefont{and}
	\bibinfo{author}{\bibfnamefont{J.~R.} \bibnamefont{Kirtley}},
	\bibinfo{journal}{Rev. Mod. Phys.} \textbf{\bibinfo{volume}{72}},
	\bibinfo{pages}{969} (\bibinfo{year}{2000}),
	\urlprefix\url{https://link.aps.org/doi/10.1103/RevModPhys.72.969}.
	
	\bibitem[{\citenamefont{Sigrist and M.~Rice}(1992)}]{Sigrist1992}
	\bibinfo{author}{\bibfnamefont{M.}~\bibnamefont{Sigrist}} \bibnamefont{and}
	\bibinfo{author}{\bibfnamefont{T.}~\bibnamefont{M.~Rice}},
	\bibinfo{journal}{Journal of the Physical Society of Japan}
	\textbf{\bibinfo{volume}{61}}, \bibinfo{pages}{4283} (\bibinfo{year}{1992}),
	\eprint{https://doi.org/10.1143/JPSJ.61.4283},
	\urlprefix\url{https://doi.org/10.1143/JPSJ.61.4283}.
	
	\bibitem[{\citenamefont{Wollman et~al.}(1993)\citenamefont{Wollman,
			Van~Harlingen, Lee, Ginsberg, and Leggett}}]{wollman_experimental_1993}
	\bibinfo{author}{\bibfnamefont{D.~A.} \bibnamefont{Wollman}},
	\bibinfo{author}{\bibfnamefont{D.~J.} \bibnamefont{Van~Harlingen}},
	\bibinfo{author}{\bibfnamefont{W.~C.} \bibnamefont{Lee}},
	\bibinfo{author}{\bibfnamefont{D.~M.} \bibnamefont{Ginsberg}},
	\bibnamefont{and} \bibinfo{author}{\bibfnamefont{A.~J.}
		\bibnamefont{Leggett}}, \bibinfo{journal}{Physical Review Letters}
	\textbf{\bibinfo{volume}{71}}, \bibinfo{pages}{2134} (\bibinfo{year}{1993}),
	ISSN \bibinfo{issn}{0031-9007},
	\urlprefix\url{https://link.aps.org/doi/10.1103/PhysRevLett.71.2134}.
	
	\bibitem[{\citenamefont{Wollman et~al.}(1995)\citenamefont{Wollman, Harlingen,
			Giapintzakis, and Ginsberg}}]{Wollman1995}
	\bibinfo{author}{\bibfnamefont{D.~A.} \bibnamefont{Wollman}},
	\bibinfo{author}{\bibfnamefont{D.~J.~V.} \bibnamefont{Harlingen}},
	\bibinfo{author}{\bibfnamefont{J.}~\bibnamefont{Giapintzakis}},
	\bibnamefont{and} \bibinfo{author}{\bibfnamefont{D.~M.}
		\bibnamefont{Ginsberg}}, \bibinfo{journal}{Physical Review Letters}
	\textbf{\bibinfo{volume}{74}}, \bibinfo{pages}{797} (\bibinfo{year}{1995}).
	
	\bibitem[{\citenamefont{Nelson et~al.}(2004)\citenamefont{Nelson, Mao, Maeno,
			and Liu}}]{nelson_odd-parity_2004}
	\bibinfo{author}{\bibfnamefont{K.~D.} \bibnamefont{Nelson}},
	\bibinfo{author}{\bibfnamefont{Z.~Q.} \bibnamefont{Mao}},
	\bibinfo{author}{\bibfnamefont{Y.}~\bibnamefont{Maeno}}, \bibnamefont{and}
	\bibinfo{author}{\bibfnamefont{Y.}~\bibnamefont{Liu}},
	\bibinfo{journal}{Science} \textbf{\bibinfo{volume}{306}},
	\bibinfo{pages}{1151} (\bibinfo{year}{2004}), ISSN \bibinfo{issn}{0036-8075,
		1095-9203},
	\urlprefix\url{http://science.sciencemag.org/content/306/5699/1151}.
	
	\bibitem[{\citenamefont{Mackenzie et~al.}(2017)\citenamefont{Mackenzie,
			Scaffidi, Hicks, and Maeno}}]{Mackenzie2017}
	\bibinfo{author}{\bibfnamefont{A.~P.} \bibnamefont{Mackenzie}},
	\bibinfo{author}{\bibfnamefont{T.}~\bibnamefont{Scaffidi}},
	\bibinfo{author}{\bibfnamefont{C.~W.} \bibnamefont{Hicks}}, \bibnamefont{and}
	\bibinfo{author}{\bibfnamefont{Y.}~\bibnamefont{Maeno}},
	\bibinfo{journal}{npj Quantum Materials} \textbf{\bibinfo{volume}{2}}
	(\bibinfo{year}{2017}).
	
	\bibitem[{\citenamefont{Li et~al.}(2019)\citenamefont{Li, Xu, Lee, Chu, and
			Chien}}]{Li2019}
	\bibinfo{author}{\bibfnamefont{Y.}~\bibnamefont{Li}},
	\bibinfo{author}{\bibfnamefont{X.}~\bibnamefont{Xu}},
	\bibinfo{author}{\bibfnamefont{M.-H.} \bibnamefont{Lee}},
	\bibinfo{author}{\bibfnamefont{M.-W.} \bibnamefont{Chu}}, \bibnamefont{and}
	\bibinfo{author}{\bibfnamefont{C.~L.} \bibnamefont{Chien}},
	\bibinfo{journal}{Science} \textbf{\bibinfo{volume}{366}},
	\bibinfo{pages}{238} (\bibinfo{year}{2019}),
	\eprint{https://www.science.org/doi/pdf/10.1126/science.aau6539},
	\urlprefix\url{https://www.science.org/doi/abs/10.1126/science.aau6539}.
	
	\bibitem[{\citenamefont{Xu et~al.}(2020)\citenamefont{Xu, Li, and
			Chien}}]{xu_spin-triplet_2020}
	\bibinfo{author}{\bibfnamefont{X.}~\bibnamefont{Xu}},
	\bibinfo{author}{\bibfnamefont{Y.}~\bibnamefont{Li}}, \bibnamefont{and}
	\bibinfo{author}{\bibfnamefont{C.}~\bibnamefont{Chien}},
	\bibinfo{journal}{Phys. Rev. Lett.} \textbf{\bibinfo{volume}{124}},
	\bibinfo{pages}{167001} (\bibinfo{year}{2020}), \bibinfo{note}{publisher:
		American Physical Society},
	\urlprefix\url{https://link.aps.org/doi/10.1103/PhysRevLett.124.167001}.
	
	\bibitem[{\citenamefont{Lv et~al.}(2017)\citenamefont{Lv, Wang, Zhang, Ding,
			Li, Wang, He, Song, Ma, and Xue}}]{Lv2017}
	\bibinfo{author}{\bibfnamefont{Y.-F.} \bibnamefont{Lv}},
	\bibinfo{author}{\bibfnamefont{W.-L.} \bibnamefont{Wang}},
	\bibinfo{author}{\bibfnamefont{Y.-M.} \bibnamefont{Zhang}},
	\bibinfo{author}{\bibfnamefont{H.}~\bibnamefont{Ding}},
	\bibinfo{author}{\bibfnamefont{W.}~\bibnamefont{Li}},
	\bibinfo{author}{\bibfnamefont{L.}~\bibnamefont{Wang}},
	\bibinfo{author}{\bibfnamefont{K.}~\bibnamefont{He}},
	\bibinfo{author}{\bibfnamefont{C.-L.} \bibnamefont{Song}},
	\bibinfo{author}{\bibfnamefont{X.-C.} \bibnamefont{Ma}}, \bibnamefont{and}
	\bibinfo{author}{\bibfnamefont{Q.-K.} \bibnamefont{Xue}},
	\bibinfo{journal}{Science Bulletin} \textbf{\bibinfo{volume}{62}},
	\bibinfo{pages}{852} (\bibinfo{year}{2017}), ISSN \bibinfo{issn}{2095-9273},
	\urlprefix\url{http://www.sciencedirect.com/science/article/pii/S2095927317302487}.
	
	\bibitem[{\citenamefont{Li et~al.}(2020)\citenamefont{Li, Xu, Lee, and
			Chien}}]{Li2020}
	\bibinfo{author}{\bibfnamefont{Y.}~\bibnamefont{Li}},
	\bibinfo{author}{\bibfnamefont{X.}~\bibnamefont{Xu}},
	\bibinfo{author}{\bibfnamefont{S.-P.} \bibnamefont{Lee}}, \bibnamefont{and}
	\bibinfo{author}{\bibfnamefont{C.~L.} \bibnamefont{Chien}},
	\bibinfo{journal}{arXiv:2003.00603 [cond-mat, physics:quant-ph]}
	(\bibinfo{year}{2020}), \bibinfo{note}{arXiv: 2003.00603},
	\urlprefix\url{http://arxiv.org/abs/2003.00603}.
	
	\bibitem[{\citenamefont{Little and Parks}(1962)}]{little_observation_1962}
	\bibinfo{author}{\bibfnamefont{W.~A.} \bibnamefont{Little}} \bibnamefont{and}
	\bibinfo{author}{\bibfnamefont{R.~D.} \bibnamefont{Parks}},
	\bibinfo{journal}{Phys. Rev. Lett.} \textbf{\bibinfo{volume}{9}},
	\bibinfo{pages}{9} (\bibinfo{year}{1962}),
	\urlprefix\url{https://link.aps.org/doi/10.1103/PhysRevLett.9.9}.
	
	\bibitem[{\citenamefont{Tinkham}(1964)}]{tinkham_consequences_1964}
	\bibinfo{author}{\bibfnamefont{M.}~\bibnamefont{Tinkham}},
	\bibinfo{journal}{Reviews of Modern Physics} \textbf{\bibinfo{volume}{36}},
	\bibinfo{pages}{268} (\bibinfo{year}{1964}), ISSN \bibinfo{issn}{0034-6861},
	\urlprefix\url{https://link.aps.org/doi/10.1103/RevModPhys.36.268}.
	
	\bibitem[{\citenamefont{Moshchalkov et~al.}(1995)\citenamefont{Moshchalkov,
			Gielen, Strunk, Jonckheere, Qiu, Haesendonck, and
			Bruynseraede}}]{Moshchalkov1995}
	\bibinfo{author}{\bibfnamefont{V.~V.} \bibnamefont{Moshchalkov}},
	\bibinfo{author}{\bibfnamefont{L.}~\bibnamefont{Gielen}},
	\bibinfo{author}{\bibfnamefont{C.}~\bibnamefont{Strunk}},
	\bibinfo{author}{\bibfnamefont{R.}~\bibnamefont{Jonckheere}},
	\bibinfo{author}{\bibfnamefont{X.}~\bibnamefont{Qiu}},
	\bibinfo{author}{\bibfnamefont{C.~V.} \bibnamefont{Haesendonck}},
	\bibnamefont{and}
	\bibinfo{author}{\bibfnamefont{Y.}~\bibnamefont{Bruynseraede}},
	\bibinfo{journal}{Nature} \textbf{\bibinfo{volume}{373}},
	\bibinfo{pages}{319} (\bibinfo{year}{1995}), ISSN \bibinfo{issn}{1476-4687},
	\urlprefix\url{https://doi.org/10.1038/373319a0}.
	
	\bibitem[{\citenamefont{Anderson and Morel}(1961)}]{ABM}
	\bibinfo{author}{\bibfnamefont{P.~W.} \bibnamefont{Anderson}} \bibnamefont{and}
	\bibinfo{author}{\bibfnamefont{P.}~\bibnamefont{Morel}},
	\bibinfo{journal}{Phys. Rev.} \textbf{\bibinfo{volume}{123}},
	\bibinfo{pages}{1911} (\bibinfo{year}{1961}),
	\urlprefix\url{https://link.aps.org/doi/10.1103/PhysRev.123.1911}.
	
	\bibitem[{\citenamefont{Balian and Werthamer}(1963)}]{BW}
	\bibinfo{author}{\bibfnamefont{R.}~\bibnamefont{Balian}} \bibnamefont{and}
	\bibinfo{author}{\bibfnamefont{N.~R.} \bibnamefont{Werthamer}},
	\bibinfo{journal}{Phys. Rev.} \textbf{\bibinfo{volume}{131}},
	\bibinfo{pages}{1553} (\bibinfo{year}{1963}),
	\urlprefix\url{https://link.aps.org/doi/10.1103/PhysRev.131.1553}.
	
	\bibitem[{\citenamefont{Mackenzie and Maeno}(2003)}]{RMP_SRO}
	\bibinfo{author}{\bibfnamefont{A.~P.} \bibnamefont{Mackenzie}}
	\bibnamefont{and} \bibinfo{author}{\bibfnamefont{Y.}~\bibnamefont{Maeno}},
	\bibinfo{journal}{Rev. Mod. Phys.} \textbf{\bibinfo{volume}{75}},
	\bibinfo{pages}{657} (\bibinfo{year}{2003}),
	\urlprefix\url{https://link.aps.org/doi/10.1103/RevModPhys.75.657}.
	
	\bibitem[{\citenamefont{Herrera et~al.}(2015)\citenamefont{Herrera,
			Guillam\'on, Galvis, Correa, Fente, Luccas, Mompean, Garc\'{\i}a-Hern\'andez,
			Vieira, Brison et~al.}}]{herrera_magnetic_2015}
	\bibinfo{author}{\bibfnamefont{E.}~\bibnamefont{Herrera}},
	\bibinfo{author}{\bibfnamefont{I.}~\bibnamefont{Guillam\'on}},
	\bibinfo{author}{\bibfnamefont{J.~A.} \bibnamefont{Galvis}},
	\bibinfo{author}{\bibfnamefont{A.}~\bibnamefont{Correa}},
	\bibinfo{author}{\bibfnamefont{A.}~\bibnamefont{Fente}},
	\bibinfo{author}{\bibfnamefont{R.~F.} \bibnamefont{Luccas}},
	\bibinfo{author}{\bibfnamefont{F.~J.} \bibnamefont{Mompean}},
	\bibinfo{author}{\bibfnamefont{M.}~\bibnamefont{Garc\'{\i}a-Hern\'andez}},
	\bibinfo{author}{\bibfnamefont{S.}~\bibnamefont{Vieira}},
	\bibinfo{author}{\bibfnamefont{J.~P.} \bibnamefont{Brison}},
	\bibnamefont{et~al.}, \bibinfo{journal}{Phys. Rev. B}
	\textbf{\bibinfo{volume}{92}}, \bibinfo{pages}{054507}
	(\bibinfo{year}{2015}),
	\urlprefix\url{https://link.aps.org/doi/10.1103/PhysRevB.92.054507}.
	
	\bibitem[{\citenamefont{Ka{\v{c}}mar{\v{c}}{\'{\i}}k
			et~al.}(2016)\citenamefont{Ka{\v{c}}mar{\v{c}}{\'{\i}}k, Pribulov{\'{a}},
			Samuely, Szab{\'{o}}, Cambel, {\v{S}}olt{\'{y}}s, Herrera, Suderow,
			Correa-Orellana, Prabhakaran et~al.}}]{kacmarcik_single-gap_2016}
	\bibinfo{author}{\bibfnamefont{J.}~\bibnamefont{Ka{\v{c}}mar{\v{c}}{\'{\i}}k}},
	\bibinfo{author}{\bibfnamefont{Z.}~\bibnamefont{Pribulov{\'{a}}}},
	\bibinfo{author}{\bibfnamefont{T.}~\bibnamefont{Samuely}},
	\bibinfo{author}{\bibfnamefont{P.}~\bibnamefont{Szab{\'{o}}}},
	\bibinfo{author}{\bibfnamefont{V.}~\bibnamefont{Cambel}},
	\bibinfo{author}{\bibfnamefont{J.}~\bibnamefont{{\v{S}}olt{\'{y}}s}},
	\bibinfo{author}{\bibfnamefont{E.}~\bibnamefont{Herrera}},
	\bibinfo{author}{\bibfnamefont{H.}~\bibnamefont{Suderow}},
	\bibinfo{author}{\bibfnamefont{A.}~\bibnamefont{Correa-Orellana}},
	\bibinfo{author}{\bibfnamefont{D.}~\bibnamefont{Prabhakaran}},
	\bibnamefont{et~al.}, \bibinfo{journal}{Phys. Rev. B}
	\textbf{\bibinfo{volume}{93}}, \bibinfo{pages}{144502}
	(\bibinfo{year}{2016}),
	\urlprefix\url{https://link.aps.org/doi/10.1103/PhysRevB.93.144502}.
	
	\bibitem[{\citenamefont{Biswas et~al.}(2016)\citenamefont{Biswas, Mazzone,
			Sibille, Pomjakushina, Conder, Luetkens, Baines, Gavilano, Kenzelmann, Amato
			et~al.}}]{muSR_fullygapped}
	\bibinfo{author}{\bibfnamefont{P.~K.} \bibnamefont{Biswas}},
	\bibinfo{author}{\bibfnamefont{D.~G.} \bibnamefont{Mazzone}},
	\bibinfo{author}{\bibfnamefont{R.}~\bibnamefont{Sibille}},
	\bibinfo{author}{\bibfnamefont{E.}~\bibnamefont{Pomjakushina}},
	\bibinfo{author}{\bibfnamefont{K.}~\bibnamefont{Conder}},
	\bibinfo{author}{\bibfnamefont{H.}~\bibnamefont{Luetkens}},
	\bibinfo{author}{\bibfnamefont{C.}~\bibnamefont{Baines}},
	\bibinfo{author}{\bibfnamefont{J.~L.} \bibnamefont{Gavilano}},
	\bibinfo{author}{\bibfnamefont{M.}~\bibnamefont{Kenzelmann}},
	\bibinfo{author}{\bibfnamefont{A.}~\bibnamefont{Amato}},
	\bibnamefont{et~al.}, \bibinfo{journal}{Phys. Rev. B}
	\textbf{\bibinfo{volume}{93}}, \bibinfo{pages}{220504}
	(\bibinfo{year}{2016}),
	\urlprefix\url{https://link.aps.org/doi/10.1103/PhysRevB.93.220504}.
	
	\bibitem[{\citenamefont{Chen et~al.}(2020)\citenamefont{Chen, Wang, Pang, Su,
			Chen, and Yuan}}]{HuiqiuNodeless}
	\bibinfo{author}{\bibfnamefont{J.}~\bibnamefont{Chen}},
	\bibinfo{author}{\bibfnamefont{A.}~\bibnamefont{Wang}},
	\bibinfo{author}{\bibfnamefont{G.}~\bibnamefont{Pang}},
	\bibinfo{author}{\bibfnamefont{H.}~\bibnamefont{Su}},
	\bibinfo{author}{\bibfnamefont{Y.}~\bibnamefont{Chen}}, \bibnamefont{and}
	\bibinfo{author}{\bibfnamefont{H.}~\bibnamefont{Yuan}},
	\bibinfo{journal}{Phys. Rev. B} \textbf{\bibinfo{volume}{101}},
	\bibinfo{pages}{054514} (\bibinfo{year}{2020}),
	\urlprefix\url{https://link.aps.org/doi/10.1103/PhysRevB.101.054514}.
	
\end{thebibliography}


%

\clearpage

\begin{figure}
	\centering
	\includegraphics[width=8cm]{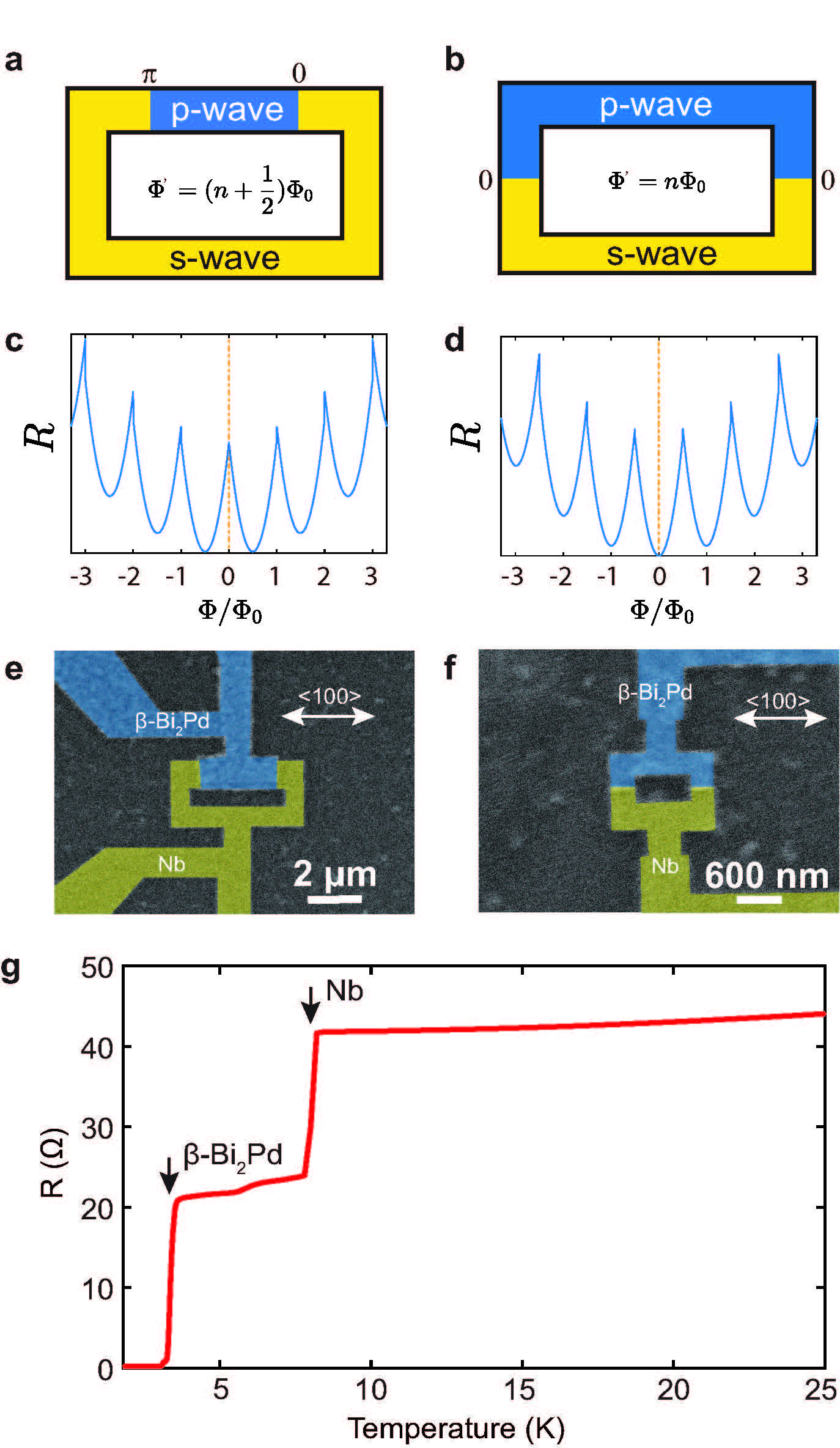}
	\caption{ \textbf{Two types of triplet/singlet composite ring structures of GLB and SS devices.}  (a) Schematic and (e) actual GLB ring of a triplet superconductor $\beta$-Bi$_2$Pd with the opposite sides connected at the two ends with s-wave Nb exhibits only half-flux $(n+1/2)\Phi_0$ (c). (b) Schematic and (f) actual SS ring of a triplet superconductor $\beta$-Bi$_2$Pd with the same side connected at two ends with s-wave Nb exhibits integer flux of n$\Phi_0$ (d).  In both cases, $\beta$-Bi$_2$Pd is an epitaxial layer of 70 nm thick oriented laterally along the [100] direction. (g) A representative resistance vs. temperature curve of the SQUID device showing the $T_c$ of $\beta$-Bi$_2$Pd and Nb. }
		
\end{figure}

\begin{figure}
	\centering
	\includegraphics[width=12cm]{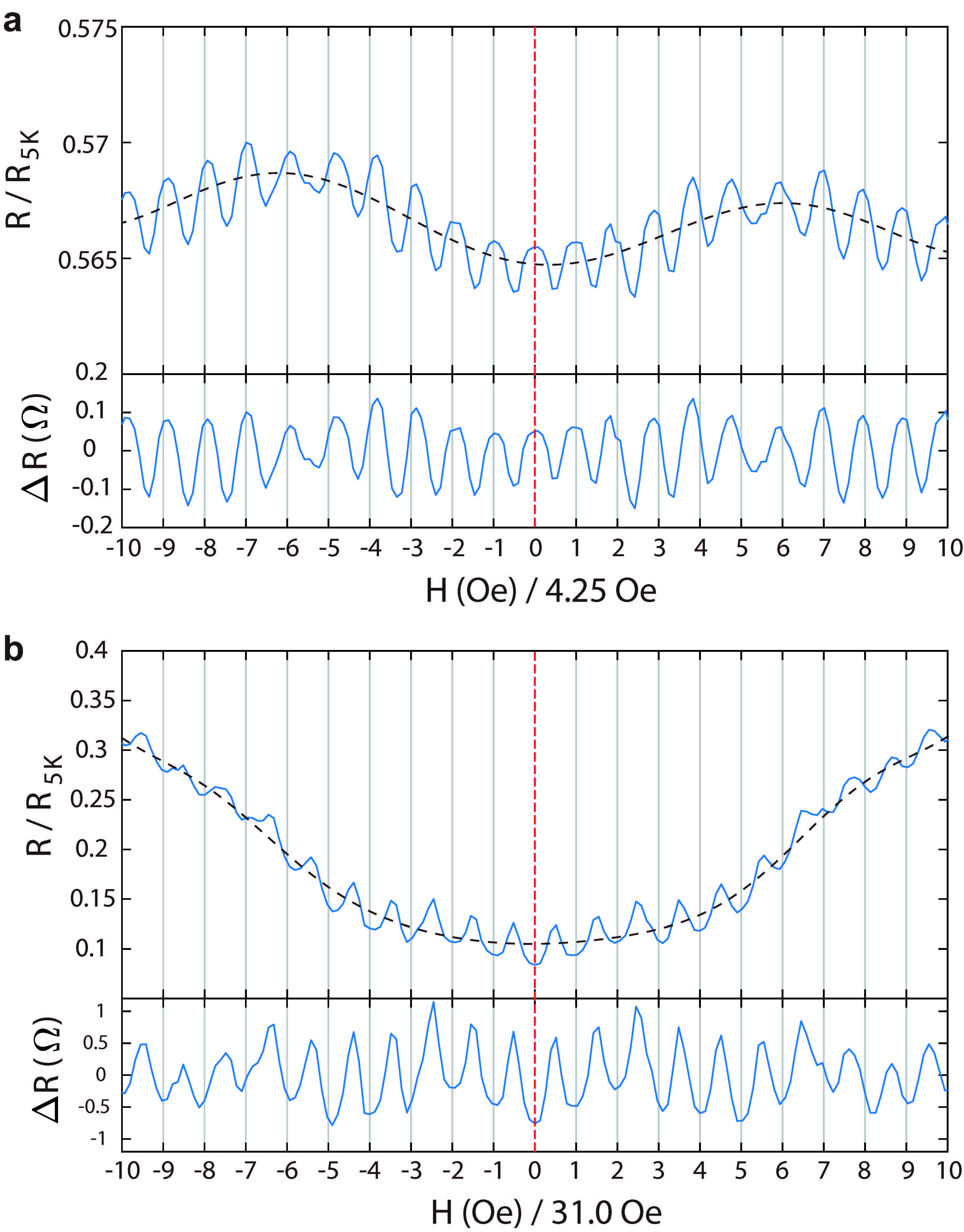}
	\caption{ \textbf{The Little-Parks effect of GLB-A and SS-B composite rings.} Normalized resistance (upper panel) including the background as the dashed curve, and the relative resistance (lower panel) with the background removed of (a) GLB-A (4.2 $\mu$m $\times$ 1.6 $\mu$m) showing the half-integer flux $(n + 1/2)\Phi_0$ and (b) SS-B (1.1 $\mu$m $\times$ 0.7 $\mu$m) showing the integer flux n$\Phi_0$, as a function of the magnetic field applied perpendicular to the ring area in units 4.25 Oe and 31.0 Oe respectively dictated by the ring size.  Vertical dashes lines at integer flux n$\Phi_0$ are indicated. }
\end{figure}

\begin{figure}
	\centering
	\includegraphics[width=16cm]{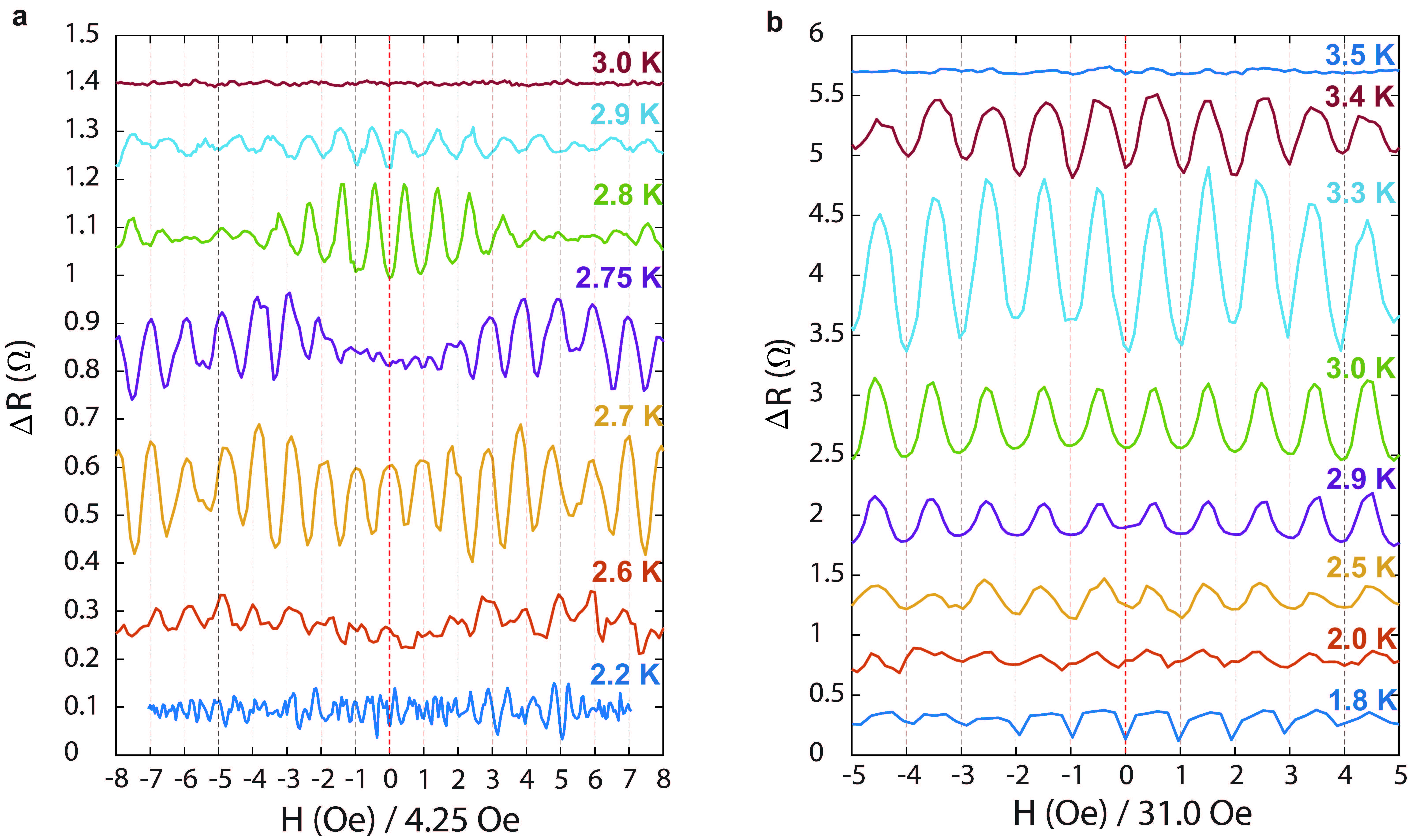}
	\caption{\textbf{Temperature dependence of the Little-Parks effect in GLB-A and SS-B rings.}  (a) GLB-A (4.2 $\mu$m $\times$ 1.6 $\mu$m) with Nb connected to the opposite ends of epitaxial $\beta$-Bi$_2$Pd segment shows half-flux at 2.6 K, integer flux at 2.9 K and transition region in between, (b) SS-B (1.1 $\mu$m $\times$ 0.7 $\mu$m) shows integer flux throughout.  Vertical dashes lines at integer flux n$\Phi_0$ are indicated. }
\end{figure}

\begin{figure}
	\centering
	\includegraphics[width=9cm]{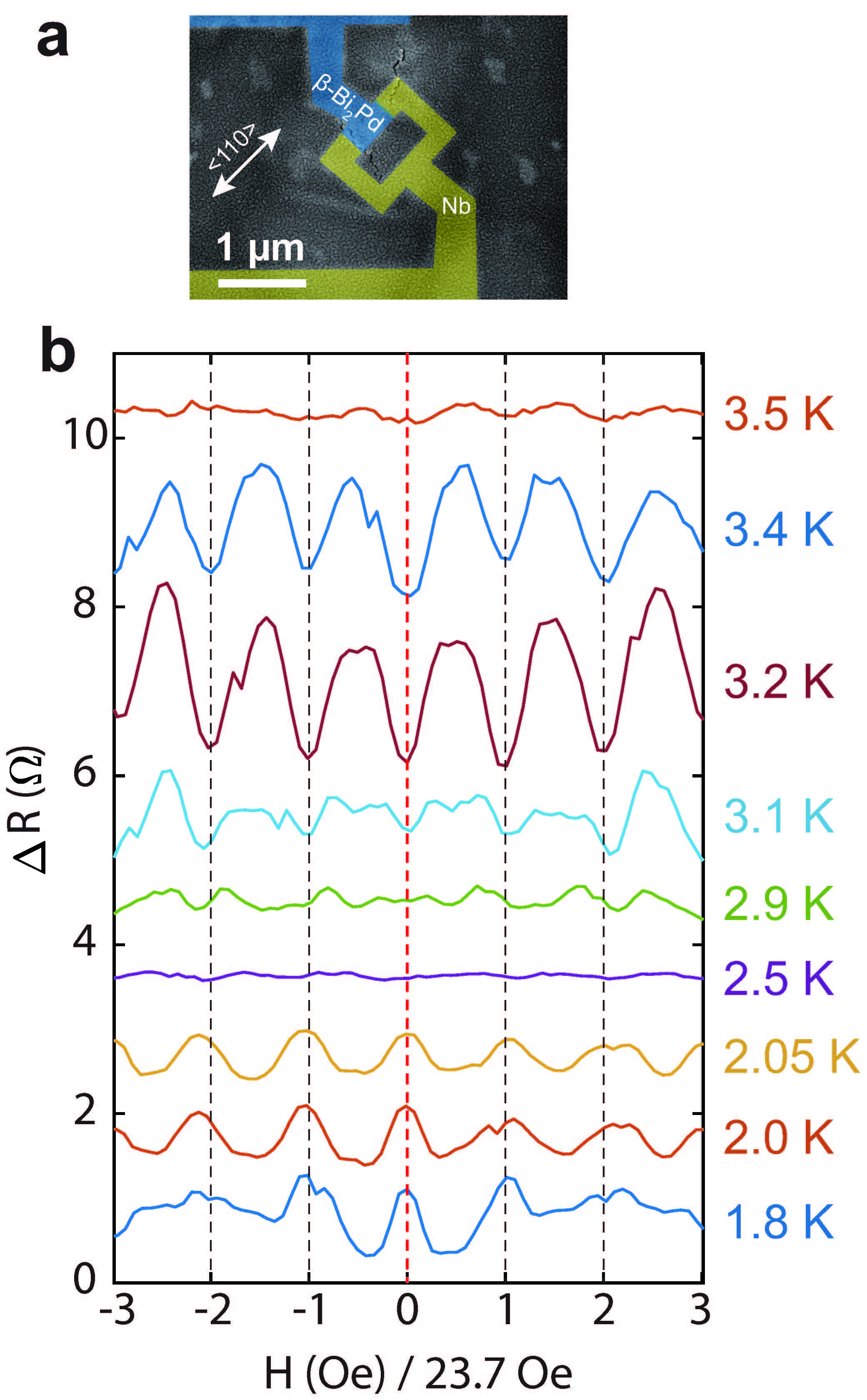}
	\caption{\textbf{Temperature dependence of the Little-Parks effect in GLB-C.}  (a) SEM image of GLB SQUID-C (1.2 $\mu$m $\times$ 0.75 $\mu$m) with Nb connected to the opposite ends of epitaxial $\beta$-Bi$_2$Pd segment along the $<110>$ direction, (b) the Little-Parks oscillation showing the transition from half-flux at 2.05 K to integer flux at 3.2 K.  Vertical dashes lines at integer flux n$\Phi_0$ are indicated. }
\end{figure}

\clearpage

\end{document}